\documentclass[aps,prl,twocolumn,showkeys,preprintnumbers]{revtex4}
\usepackage{graphicx,amsmath}
\usepackage{dcolumn}
\usepackage{amsmath}
\usepackage{amssymb}
\usepackage{amsfonts}

\begin{document}

\title{Surfaces of Constant Temperature in Time}

\author{David~Ford}
\email[]{dkf0rd@netscape.net}
\affiliation{Department of Physics, Naval Postgraduate School, Monterey, California}

\date{\today}

\begin{abstract}
The inverse relationship between energy and time is as familiar as Planck's constant.
From the point of view of a system with many states, perhaps a better representation of
the system is a vector of characteristic
times (one per state) for example, in the case of a canonically distributed system. In the vector case
the inverse relationship persists, this time as a relation between the $L_2$ norms. That relationship
is derived herein. An unexpected
benefit of the vectorized time viewpoint is the determination of surfaces of constant temperature in terms of the 
time coordinates. The results apply to all empirically accessible systems, that is situations where
details of the dynamics are recorded at the microscopic level of detail. This includes all manner
of simulation data of statistical mechanical systems as well as experimental data from
actual systems (e.g. the internet, financial market data) where statistical physical methods 
have been applied.
\end{abstract}


\maketitle

\title{Surfaces of Constant Temperature in Time}         
\author{David Ford\\ Naval Postgraduate School\\ Monterey, California }        

\section{Introduction}
Perhaps the most obvious relation between energy and time is given  by the expression for
the energy of a single photon $E= \hbar \nu$.  In higher dimensions
a similar energy-time relation
$$
\| H \|_2 =\frac{const.}{\| t \|_2}
$$
\noindent holds for the $L_2$ norms of state energies and characteristic times associated with a canonically distributed system. This relationship is made precise herein. A by-product of the result is the possibility of the determination of surfaces of constant temperature given sufficient details about the trajectory of the system through its path space.

As an initial value problem, system kinetics are determined once an initial state and energy function are specified.
In the classical setting, representative cell occupation numbers may be assigned any compact region of  position, 
$\mathbf{q}$, momentum, $\mathbf{p}$, space \cite{uhl}. An important model of a quantum system is provided by lattices  
of the Ising type \cite{glauber}, \cite{mnv}. Here the state of the system is typically specified by the configuration of spins.

Importantly, two systems that share exactly the same state space may assign energy levels to those states differently.
In the classical context one may, for example,  hold the momenta fixed and vary the energy by preparing a second system of slower moving but more massive particles. In the lattice example one might compare systems with the same number of sites but different coupling constants, etc.

Consider a single large system comprised of an ensemble 
of smaller subsystems  which all
share a common, finite state space. Let the state energy assignments vary from one 
subsystem  to the next.
Equivalently, one could consider a single, fixed member of the ensemble whose 
Hamiltonian, $H_{subsystem}$, is somehow varied (perhaps by varying 
external fields, potentials and the like \cite{schro}).    
Two observers, A and B, monitoring two different members of the ensemble, $\mathcal{E}_A$
and  $\mathcal{E}_B$, would accumulate the same lists of states visited 
but different lists of state occupation times. 
The totality of these characteristic time scales, when interpreted as a list of coordinates (one list per member
of the ensemble),  sketch out a surface of constant temperature in the shared coordinate space.

\section{Restriction on the Variations of $H_{subsystem}$ }

From the point of view of simple arithmetic, any variation of $H_{subsystem}$ 
is permissible but recall there are constraints inherent in the construction of a
canonical ensemble. Once an energy reference for the subsystem has been declared,
the addition of a single constant energy  uniformly to all subsystems states will not be allowed. 

Translations of $H_{subsystem}$ are temperature changes in the bath.
The trajectory of the \text { \it total } system takes place in a thin energy shell. If the fluctuations of the subsystem are shifted uniformly then the fluctuations in the bath are also shifted uniformly (in the opposite direction).
This constitutes a change in temperature of the system.  This seemingly banal observation is not
without its implications. The particulars of the situation are not unfamiliar.

A similar concept from Newtonian mechanics is the idea of describing the motion
of a system of point masses  from the frame of reference of the mass center.
Let $\{ H_1, H_2, \ldots , H_N \}$ be the energies of an $N-$state system. 
A different Hamiltonian
might assign energies to those same states differently, say  $\{ \tilde H_1, \tilde H_2, \ldots , \tilde H_N \}$.
To describe the transition from the energy assignment $ \mathbf{H}$ to the assignment  $\mathbf{\tilde H}$
one might first rearrange the values about the original `mass center' 
\begin{equation}
\frac{ H_1+ H_2+ \ldots + H_N}{N}
\end{equation}
and then uniformly shift the entire assembly
to the new `mass center'
\begin{equation}
\frac{  \tilde H_1+  \tilde H_2+ \ldots +  \tilde H_N}{N}.
\end{equation} 
In the present context, the uniform translations of the subsystem state energies
are temperature changes in the bath.
As a result, the following convention is adopted.  
 For a given set of state energies
 $\{ H_1, H_2, \ldots , H_N \}$,
 only those changes to the state energy assignments that 
 leave the `mass center'
\noindent unchanged will be considered in the sequel. 

The  fixed energy value of the ``mass center'' serves
as a reference energy in what follows. For simplicity this reference is taken to be zero.
That is
\begin{equation}\label{zero}
 H_1+ H_2+ \ldots + H_N = 0.
\end{equation}
\noindent  Uniform translation will be treated as a temperature fluctuation in what follows.
An obvious consequence is that only $N-1$ subsystem state energies and the bath temperature
$\theta$ are required to describe the statistics of a canonically distributed system.

\section{Two One-Dimensional Subspaces }
In the event that a trajectory of the subsystem is observed
long enough so that each of the $N$ states
  is visited many times,  it is supposed  that the vector of occupancy times spent in state,
 $\{ \Delta t_1, \Delta t_2, \ldots, \Delta t_N \}$, is connected to any vector of N-1 independent
 state energies and the common bath temperature, $\{ H_1, H_2, \ldots , H_{N-1}, \theta \}$, 
 by relations of the form 
\begin{equation}\label{t&e_ratio1} 
\frac{ \Delta t_{k} }{  \Delta t_{j}  }=\frac{ e^{- \frac{H_{k}}{\theta}} }{e^{- \frac{H_{j}}{\theta}}}
\end{equation}
\noindent for any $k, j \in \{1,2,\ldots,N\}$. The value of the omitted state energy, $H_N$, is determined by
equation (\ref{zero}).

The number of discrete visits to at least one of these states will 
be a minimum. Select one of these minimally visited states and label it the rare state.
The observed trajectory may be decomposed into cycles beginning and ending on visits to the rare state and
the statistics of a typical cycle may be computed. For each $k \in \{1,2,\ldots,N\}$,
let  $\Delta t_k$ represent the amount of continuous time spent in the $k^{th}$ state during a typical cycle.
In the Markoff setting the $L_1$ norm
 \begin{equation}\label{cd} 
\sum_{k=1}^N  \Delta t_{k} = \textrm{characteristic system time},
\end{equation}
\noindent may serve as the Carlson depth. These agreements do not affect the validity of
equation (\ref{t&e_ratio1}).

At finite temperature, it may be the case that the system is uniformly distributed. 
That is,  the observed subsystem trajectory is representative
of the limiting case where the interaction Hamiltonian has  been turned off 
and the subsystem dynamics take place on a surface of constant energy.

\begin{figure}[htbp]
\begin{center}
\leavevmode
\includegraphics[width=60mm,keepaspectratio]{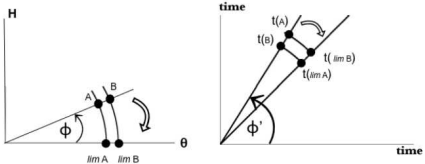}
\caption{Schematic of the approach to the uniform distribution for dilatation pairs in
both the energy and time domains.}
\label{CLscale}
\end{center}
\end{figure}

In the left hand panel of figure \ref{CLscale}, the $\theta-$axis coincides with the set of all state energies and 
bath temperatures
corresponding to uniformly distributed systems. 
In the time domain, the ray containing the vector $\mathbf{1}$ (see the right hand panel) depicts the set of state occupancy times that give rise to uniformly distributed systems.

For real constants $c_{\Delta t}$ and $c_{E}$ scale transformations of the type 
\begin{eqnarray*}
\Delta \mathbf{t}             \longrightarrow    &c_{\Delta t} \; \; \Delta \mathbf{t} \\
\{ \mathbf{H}, \theta \}    \longrightarrow   &c_{E} \; \{ \mathbf{H}, \theta \}
\end{eqnarray*}
\noindent  dilatate points along rays in their respective spaces and leave equation (\ref{t&e_ratio1}) invariant.

The left hand panel of  figure \ref{CLscale} shows a pair of energy, temperature coordinates:  A and B,
related by a dilatation scale factor $c_{E}$, rotated successively toward the coordinates $lim \, A$ and $lim \, B$ 
which lie on the line of uniform distribution (the $\theta$ axis) in the energy, temperature domain. Throughout the limit process (parameterized by the angle $\phi$) the scale factor $ c_{E}$ is held constant. 
Consistent with the relations in equation (\ref{t&e_ratio1}),  the points  $t(A) $ and $t(B)$ (putative time domain images of the given energy, temperature domain points A and B) as well as the image of their approach to the
uniform distribution in  time  ($\phi ' = cos^{-1}(\frac{1}{\sqrt{N}}) $, where N is the dimensionality of the system), are shown in the right hand panel of the same figure. 

As the angle of rotation $\phi'$ (the putative image of $\phi$ in the time domain) is varied,  there is the possibility of a consequent variation of the time domain dilatation scale factor  $c_{\Delta t}$ that maps $t(A) $ into $t(B)$. That is,
 $c_{\Delta t}$ is an unknown function of $\phi'$. However in the limit of zero
interaction between the subsystem and the bath 
the unknown time domain scaling, $c_{\Delta t}$, consistent with the given energy, temperature
scaling, $ c_{E}$, is rather easily obtained. 

At any step in the limit process as $\phi' $ approaches $cos^{-1}(\frac{1}{\sqrt{N}})$ equation (\ref{t&e_ratio1}) implies that
\begin{equation}\label{t&e_ratio2} 
\frac{ \Delta t(B)_{k} }{  \Delta t(B)_{j}  }=\frac{ \Delta t(A)_{k} }{  \Delta t(A)_{j}  }
\end{equation}
\noindent for any $k, j \in \{1,2,\ldots,N\}$.  

Assuming, as the subsystem transitions from  weakly interacting to  conservative, that there are no discontinuities
in the dynamics, then equations  (\ref{t&e_ratio1}) and (\ref{t&e_ratio2})
hold along the center line $\phi ' = cos^{-1}(\frac{1}{\sqrt{N}})$ as well. 

\begin{figure}[htbp]
\begin{center}
\leavevmode
\includegraphics[width=60mm,keepaspectratio]{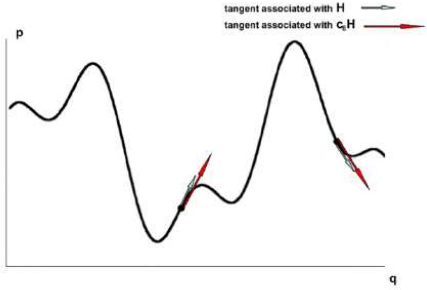}
\caption{The trajectory is everywhere tangent to both $\mathbf{H}$ and  $\mathbf{c_E H}$ vector fields.}
\label{trajectory}
\end{center}
\end{figure}

In the conservative case with constant energy $H_{ref}$, the set identity 
\begin{widetext}
\begin{equation}\label{setidentity} 
 \{ (\mathbf{q},\mathbf{p}): \mathbf{H}(\mathbf{q},\mathbf{p}) - H_{ref} = 0 \} \equiv 
\{ (\mathbf{q},\mathbf{p}): c_{E} \; ( \mathbf{H}(\mathbf{q},\mathbf{p}) -  H_{ref} ) = 0  \}
\end{equation} 
\end{widetext}

\noindent together with scaling behavior of the position and momentum velocities given by Hamilton's equations
\begin{equation}\label{spedup}
\begin{split}
 \mathbf{ \dot{q}(A)} \rightarrow & c_{E} \, \mathbf{ \dot{q}(A)} \\
 \mathbf{ \dot{p}(A)} \rightarrow & c_{E} \, \mathbf{ \dot{p}(A)}
 \end{split}
 \end{equation}

\noindent illustrate that the phase space trajectory associated with the energy, temperature domain point $lim B$  is simply the trajectory at the point $lim A$ with a time parameterization ``sped up'' by the scale factor $c_{E}$. See figure \ref{trajectory}. 
This identifies the the scale factor associated with
the points $t(lim B)$ and $t(lim A)$ as

\begin{equation}\label{limCt} 
\lim_{\phi ' \rightarrow cos^{-1}(\frac{1}{\sqrt{N}})} c_{\Delta t}(\phi') = \frac{1}{c_E}.
\end{equation}

\section{ Matched Invariants Principle and the Derivation of the Temperature Formula}

A single experiment is performed and two observers are present.
The output of the single experiment is two data points (one per observer): a single point in 
in the $\Delta t$ space and a single point in the $(H;\theta)$ space.

In the event that another  experiment is performed and the observers repeat the
activity of the previous paragraph, the data points generated are either both the same
as the ones they produced as a result of the first experiment
or else both are different. If a series of experiments are under observation,
after many iterations the sequence of data points generated traces out a curve.
There will be one curve in each space.  

$\it{The \; principle \; follows}$: in terms of probabilites, the two observers will 
produce consistent results in the case when the data points
(in their respective spaces) have changed from the first experiment to the second
but the probabilites have not. That is, if one observer experiences a dilatation
so does the other.

Of course, if the observers are able to agree if dilatation has occurred they are also able to agree
that it has not.
In terms of probability gradients, in either space the dilatation
direction is the direction in which all the probabilities are invariant.
In the setting of a system with N possible states,
the N-1 dimensional space perp to the dilatation is spanned by 
any  set of N-1 probability gradients.  We turn next to an application
of the MIP.

\begin{figure}[htbp]
\begin{center}
\leavevmode
\includegraphics[width=60mm,keepaspectratio]{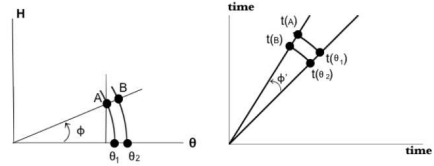}
\caption{The temperature ratio is invariant with respect to rotation in either space }
\label{arcs}
\end{center}
\end{figure}

Consider two points $\theta_1$ and $\theta_2$ along a ray colocated with the temperature axis in the $(H,\theta)$ space.
Suppose that the ray undergoes a rigid rotation (no dilatation) and that in this way the two points are
mapped to two new points $A$ and $B$ along a ray which makes an angle $\phi$ with the temperature axis.
See the left hand panel of figure \ref{arcs}.

It's pretty obvious that the temperature ratio is preserved throughout the motion. For whatever the angle
$\phi$  

\begin{equation}
\frac{ \theta_1}{   \theta_2  }=\frac{ \theta_1 \;cos(\phi)}{   \theta_2 \;cos(\phi) }=\frac{ \theta(A)}{   \theta(B) }.
\end{equation}

Let $t(\theta_1)$ and $t(\theta_2)$ be the images in the time domain of the points $\theta_1$ and
$\theta_2$ in $(H, \theta)$ space. According to the matched invariants principle, since the rotation 
in $(H, \theta)$ space was rigid so the corresponding motion as mapped to the  time domain is also a rigid rotation
(no dilatations). See figure \ref{arcs}.

More precisely, to the generic point $A$ in $(H, \theta)$ space with coordinates $(H_1,H_2,\ldots,H_{N}, \theta)$
associate a magnitude, denoted $\| \mathbf{H}\|$, and a unit vector $\hat{ \mathbf{e}}_{\mathbf{H}}$. 
Recall that the $H$'s live on the hyperplane $H_1 + H_2 + \cdots + H_N =0.$
It will be convenient to express the unit vector in the form

\begin{equation}
\hat{ \mathbf{e}}_{\mathbf{H}} =\frac{ \{\frac{H_{1}}{\theta},\frac{H_{2}}{\theta},\ldots,\frac{H_{N}}{\theta},1\} }
{  \sqrt{ (\frac{H_1}{\theta})^2+(\frac{H_2}{\theta})^2+\cdots+(\frac{H_{N}}{\theta})^2+1        }}.
\end{equation}

The angle between that unit vector and the temperature axis is determined by
\begin{equation}
cos(\phi) = \hat{ \mathbf{e}}_{\theta} \cdot \hat{ \mathbf{e}}_{\mathbf{H}} 
\end{equation}

\noindent where $\hat{ \mathbf{e}}_{\theta} = \{0,0,\ldots,0,1\}$.

The temperature at the point $A$, is the projection of its magnitude, $\| \mathbf{H}_A\|$, onto
the temperature axis

\begin{equation}
\theta(A)= \| \mathbf{H}_A\| \,cos(\phi).
\end{equation}

Another interpretation of the magnitude $\| \mathbf{H}_A\|$ is as the temperature at the point  $\theta_1$,
the image of $A$ under a rigid rotation of the ray containing it,
on the temperature axis. See figure \ref{arcs}. With this interpretation

\begin{equation}\label{punchline}
\theta(A)= \theta_1 \,cos(\phi).
\end{equation}
An easy consequence of equation (\ref{zero}) is
\begin{equation}\label{firstformula}
\frac{H_k}{\theta} =  \log [ \frac{( \prod_{j=1}^N p_j )^{\frac{1}{N}}}{p_k} ].
\end{equation}
In terms of the occupation times
\begin{equation}\label{firstformulaA}
\frac{H_k}{\theta} =  \log [ \frac{( \prod_{j=1}^N \Delta t_j )^{\frac{1}{N}}}{\Delta t_k} ].
\end{equation}
An easy implication of equation (\ref {limCt}) is that
\begin{equation}\label{centerline}
\sqrt{\sum_{j=1}^N \Delta t_j^2}=  \frac{\textrm{const.}}{\theta_1}.
\end{equation}
\noindent for an arbitrary but fixed constant carrying dimensions of $\textrm{time}\cdot\textrm{energy}$.

Together equations (\ref{punchline}), (\ref{firstformulaA}), and (\ref{centerline}) 
uniquely specify the surfaces of constant temperature in time
\begin{figure}[htbp]
\begin{center}
\leavevmode
\includegraphics[width=60mm,keepaspectratio]{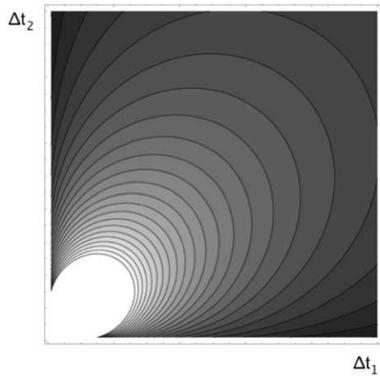}
\caption{The constant temperature surfaces for a two dimensional system. }
\label{contours}
\end{center}
\end{figure}
\begin{widetext}
\begin{equation}\label{daformula}
\theta(\Delta \mathbf{t})= \frac{ \textrm{const.} }
{ \| t \|_2 \,  \sqrt{ (\log [ \frac{ \prod }{\Delta t_1} ])^2+(\log [ \frac{ \prod }{\Delta t_2} ])^2+\cdots+
(\log [ \frac{ \prod }{\Delta t_{N}} ])^2+1        }}
\end{equation}
\end{widetext}
\noindent where,
\begin{equation}
\prod = (\Delta t_1\cdot \Delta t_2 \ldots \Delta t_N)^{\frac{1}{N}}.
\end{equation}
The temperature formula (\ref{daformula}) may be recast into the more familiar form
\begin{equation}
\| H \|_2 =\frac{const.}{\| t \|_2}
\end{equation}
With the temperature determined, equation (\ref{firstformulaA}) gives the state energies of a canonically 
distributed subsystem. From these, a wealth of useful macroscopic properties of the dynamics may be computed \cite{Fo2}. Surfaces of constant temperature for a two state system are shown in figure 4.

\bibliography{theta_aps_bib}

\end{document}